\documentclass[12pt]{article}
\usepackage{amsmath}
\usepackage{amssymb}
\usepackage{graphicx}
\usepackage{epsfig}
\usepackage{color}
\usepackage{fullpage}

\newcommand\pubdate{ }

\def\MITMKI{Dept. of Physics,\\
 Massachusetts Institute of Technology, Cambridge, MA 02139, USA}
 \def\MITMATH{Dept. of Mathematics\\
 Massachusetts Institute of Technology , Cambridge, MA 02139, USA}

\def\Title#1{\begin{center} {\Large #1 } \end{center}}
\def\Author#1{\begin{center}{ \sc #1} \end{center}}
\def\Address#1{\begin{center}{ \it #1} \end{center}}

\def\andauth{\begin{center}{and} \end{center}}
\newcommand\pubblock{\rightline{\begin{tabular}{l}\\
         \pubdate \end{tabular}}}

\title{Oscillons in the early universe}
\author{Mustafa A. Amin \& David Shirokoff}

\date{ }
\begin{document}

\begin{titlepage}
\pubblock

\vfill \Title{Flat-top oscillons in an expanding universe}
\vfill \Author{Mustafa A. Amin\footnote{email: mamin@mit.edu}} \Address{\MITMKI} \andauth
\Author{David Shirokoff\footnote{email: shirokof@mit.edu}}\Address{\MITMATH}

\begin{abstract}
Oscillons are extremely long-lived, oscillatory, spatially localized field configurations that arise from generic initial conditions in a large number of nonlinear field theories. With an eye towards their cosmological implications, we investigate their properties in an expanding universe. We (1) provide an analytic solution for one-dimensional oscillons (for the models under consideration) and discuss their generalization to three dimensions, (2) discuss their stability against long wavelength perturbations,  and (3) estimate the effects of expansion on their shapes and lifetimes. In particular, we discuss a new, extended class of oscillons with surprisingly flat tops. We show that these flat-topped oscillons are more robust against collapse instabilities in (3+1) dimensions than their usual counterparts. Unlike the solutions found in the small amplitude analysis, the width of these configurations is a nonmonotonic function of their amplitudes.

\end{abstract}
\vfill
\end{titlepage}
\def\thefootnote{\fnsymbol{footnote}}
\setcounter{footnote}{0}

\section{Introduction}
\label{Intro}
A number of physical phenomenon from water waves traveling in narrow canals \cite{Russell:1844}, to phase transitions in the early Universe \cite{Frieman:1988ut} exhibit the formation of localized,  energy density configurations, even without gravitational interactions. The reason for their longevity are varied. Some configurations are stable due to conservation of topological or nontopological charges, while some are  due to a dynamical balance between the  nonlinearities and dissipative forces. 

Relativistic, scalar field theories (with nonlinear potentials) form simple yet interesting candidates for studying such phenomenon. Some well-studied examples include topological solitons in the $1+1$-dimensional Sine-Gordon model and nontopological solitons such as Q-balls \cite{Coleman:1985ki}. The Sine-Gordon soliton is stationary in time whereas the Q-balls are oscillatory in nature. Both have conserved charges which make them stable (at least without coupling to gravity). This paper deals with another interesting example of such localized configurations called oscillons (also called breathers). Like the Sine-Gordon soliton, they can exist in real scalar fields, and like the Q-balls they are oscillatory in nature. Unlike both of the above examples they do not have any known conserved charges (however, see \cite{Kasuya:2002zs} for an adiabatic invariant). In general they decay, however their lifetimes are significantly longer than any natural time scales present in the lagrangian. Along with their longevity, another fascinating aspect of oscillons is that they emerge naturally from relatively arbitrary initial conditions.

Not all scalar field theories support oscillons. In the next section we discuss the requirements for the potential. Here, we note that the requirement is satisfied by a large number of physically well-motivated examples. For example, the potential for the axion, as well as almost any potential near a vacuum expectation value related to symmetry breaking, support oscillons. Oscillons have also been found in the restricted standard model $SU(2)\times U(1)$, \cite{Farhi:2005rz, Graham:2006vy, Graham:2006xs}. 

Oscillons first made their appearance in the literature in the 1970s \cite{Bogolyubsky:1976yu}. They were subsequently rediscovered in the 1990s \cite{Gleiser:1993pt}. Oscillons are not exact solutions and (very slowly) radiate their energy away. The amplitude of the outgoing radiation (in the small amplitude expansion) has been calculated by a number of authors, see for example \cite{Segur:1987mg, Fodor:2008es, Fodor:2009kf}. Characterization of their lifetimes and related properties using the ``Gaussian" ansatz for the spatial profile was done in \cite{Gleiser:2009ys} (also see references therein). The importance of the dimensionality of space for these objects has been discussed in \cite{Saffin:2006yk,Gleiser:2004an}. 

Their possible applications in early Universe physics has not gone unnoticed. For example, they could be relevant for axion dynamics near the QCD phase transition \cite{Kolb:1993hw}. The properties of oscillons in a $1+1$-dimensional expanding universe (in the small amplitude limit) have been discussed in \cite{Farhi:2007wj, Graham:2006xs}. Their importance during bubble collisions and phase transitions have been discussed in \cite{Dymnikova:2000dy}. In \cite{Hindmarsh:2007jb}, interactions of oscillons with each other and with domain walls were studied in 2+1 dimensions.

In this paper, we point out what is required of scalar field potentials to support oscillons. 
We then derive the frequency as well as the spatial profile of the oscillons for a class of models under consideration. We show that the spatial profile can be very different from a Gaussian, an ansatz often made in the literature. In particular, we derive the nonmonotonic relationship between the height and the width of the oscillons, and discuss the importance of this feature for the stability of oscillons (see \cite{Lee:1991ax} for a somewhat related analysis for Q-balls). To the best of our knowledge, this has not been done previously in the literature in the context of oscillons. We consider the stability of oscillons against small perturbations, mainly with spatial variations comparable to the width of the oscillons. We also comment on a possible, narrow band instability at higher wave numbers. 

Oscillons could have important applications in cosmology, especially in the early Universe.  With this in mind, we discuss the changes in the profile and  the loss of energy from these oscillons due to expansion. 

The properties of oscillons can depend significantly on the number of spatial dimensions. In this paper, for simplicity we always start with $1+1$-dimensional scenarios where analytic treatment is often possible. We then extend our results to the physically more interesting case of $3+1$ dimensions, analytically where possible and numerically otherwise. We extend previous analysis to the interesting ``flat-top" oscillons because of our new systematic method for capturing the entire range of possible amplitudes, while still using the methods from the small amplitude expansion. Our expansion, which can also be thought of as a single frequency approximation, allows us to use results existing in the literature for time periodic, localized solutions.

Although interesting as classical solutions, a quantum treatment can lead to changes in oscillon lifetimes \cite{Hertzberg:2009}. Another question worth investigating is the stability of oscillons coupled to other fields. These two questions are beyond the scope of this paper. 
 
The rest of the paper is organized as follows:  In Sec. \ref{ScalarOsc} we give a brief overview of oscillons in scalar field theories. In Sec. \ref{model} we introduce a simple model that is used throughout the paper. Section \ref{profile} deals with the derivation of the shape and frequency of the oscillons in the absence of expansion. Section \ref{stability} focuses on linear stability of oscillons. Section \ref{expansion} discusses the effects of expansion. Our conclusions and future directions are presented in Sec. \ref{con}.

\section{A gentle introduction to oscillons in scalar fields}
\label{ScalarOsc}
Oscillons are extremely long-lived, oscillatory, spatially localized field configurations that exist in a large number of nonlinear scalar field theories. We find it convenient to visualize an oscillon as a spatially localized, smooth envelope of the field value oscillating with a constant frequency. To get a heuristic understanding of what kind of the potentials support oscillons, let us consider the equation of motion for a $1+1$-dimensional scalar field: 
  \begin{equation}
 \begin{aligned}
 \Box \varphi-V'(\varphi)=0,\\
 \partial_{t}^2\varphi-\partial_x^2\varphi+V'(\varphi)=0.
 \end{aligned}
 \end{equation}
where $V'(\varphi)\rightarrow m^2\varphi$ as $\varphi\rightarrow0$. Let us approximate the oscillon field configuration as $\varphi(t,x)\sim \Phi(x)\cos[\omega t]$. For a localized configuration, as we move far enough away from the center (whereby the nonlinearity in the potential is irrelevant), we get
  \begin{equation}
 \begin{aligned}
-\omega^2\Phi-\partial_x^2\Phi+m^2\Phi\sim0.
 \end{aligned}
 \end{equation}
Again, because we are looking for a smooth, localized configuration, we must have $\omega^2< m^2$. For regions near the center of the configuration, we expect $\partial_x^2\Phi<0$ for the lowest energy solutions. This means that for the equation to be satisfied, $$V'(\Phi)-m^2\Phi<0.$$ 
Thus for oscillons to exist in potentials with a quadratic minimum, we require $V'(\varphi)< m^2\varphi$ for some range of the field value.  

It is not too difficult to think of physically motivated potentials satisfying this requirement. For example, the potential for the QCD axions $V(\varphi)=m^2f^2\left[1-\cos (\varphi/f)\right]$ where $f$ is the Peccei-Quinn scale and $m$ is the mass, or any symmetry breaking potential expanded about it's vacuum expectation value. Both potentials ``open up" a little when we move away from the minimum. 

The above (heuristic) argument does not provide a reason for the longevity of oscillons. For oscillons, their shape, which determines their Fourier content, guarantees that the amplitude at the wave number of the outgoing radiation is exponentially suppressed (at least for the small amplitude oscillons). For details see \cite{Hertzberg:2009} and the subsection on radiation in this paper. 

\section{The model}
\label{model}
We begin with the action for a real scalar field in a $d+1$-dimensional, spatially flat, homogeneous, expanding universe ($\hbar=c=1$):
\begin{equation}
{S}_{d+1}=\int (adx)^ddt \left[\frac{1}{2}\left(\partial_t\varphi\right)^2-\frac{1}{2a^2}\left(\nabla\varphi\right)^2-V(\varphi)\right],
\end{equation}
where 
\begin{equation}
\label{eq:potential}
V(\varphi)=\frac{1}{2}m^2\varphi^2-\frac{\lambda}{4}\varphi^4+\frac{g}{6}\varphi^6,
\end{equation} 
$a(t)$ is the dimensionless scale factor and $\lambda, g>0$. As we have discussed, what is crucial for the existence of oscillons is $$V'(\varphi)-m^2{\varphi}<0,$$ for some range of the field. The potential [Eq. (\ref{eq:potential})] is the simplest model which captures the effect we wish to explore, namely, a nonmonotonic relationship between the height and width and its implication for stability. However, apart from detailed expressions, our results are general and not restricted to the particular shape of the potential.

We find it convenient to work with dimensionless space-time variables as well as fields. Using $(t,x)\rightarrow m^{-1}(t,x)$ and $\varphi\rightarrow m\lambda^{-1/2}\varphi$ as well as $g\rightarrow(\lambda/m)^2g$ the action becomes
\begin{equation}
{S}_{d+1}=m^{3-d}\lambda^{-1}\int (a dx)^{d}dt \left[\frac{1}{2}\left(\partial_t\varphi\right)^2-\frac{1}{2a^2}\left(\nabla\varphi\right)^2-V(\varphi)\right],
\end{equation}
with $$V(\varphi)=\frac{1}{2}\varphi^2-\frac{1}{4}\varphi^4+\frac{g}{6}\varphi^6.$$
The classical equations of motion are given by
  \begin{equation}
 \begin{aligned}
 \partial_{t}^2\varphi-\frac{\nabla^2}{a^2}\varphi+H \partial_{t}\varphi+\varphi-\varphi^3+g\varphi^5=0,
 \end{aligned}
 \end{equation}
 where $g$  is the only free parameter in the potential and $H=\dot{a}/a$. 
We will be concentrate on the case where $g\gg1$. This gives a controlled expansion in powers of $g^{-1/2}$, which allows us to derive an analytic form for the profile. Our approach is similar to the small amplitude expansion, with the important difference that it captures the entire range of amplitudes for which oscillons exist. Moreover, in our analysis we show that the flat-top oscillons are stable against small amplitude, long wavelength perturbations on time scales of order $g$.

  \section{Oscillon profile and frequency}
  \label{profile}
  In this section we derive the spatial profile of the oscillons in our model in a $1+1$- and $3+1$-dimensional Minkowski universe. We include the effects of expansion in Sec. \ref{expansion}. 
  For simplicity, we begin with the $1+1$-dimensional case.
\subsection{Profile and frequency in $1+1$ dimensions}  
\noindent The equation of motion is
 \begin{equation}
 \partial_t^2{\varphi}-\partial_x^2\varphi+\varphi-\varphi^3+g\varphi^5=0.
 \end{equation}

\begin{figure}[t] 
   \centering
   \includegraphics[width=5in]{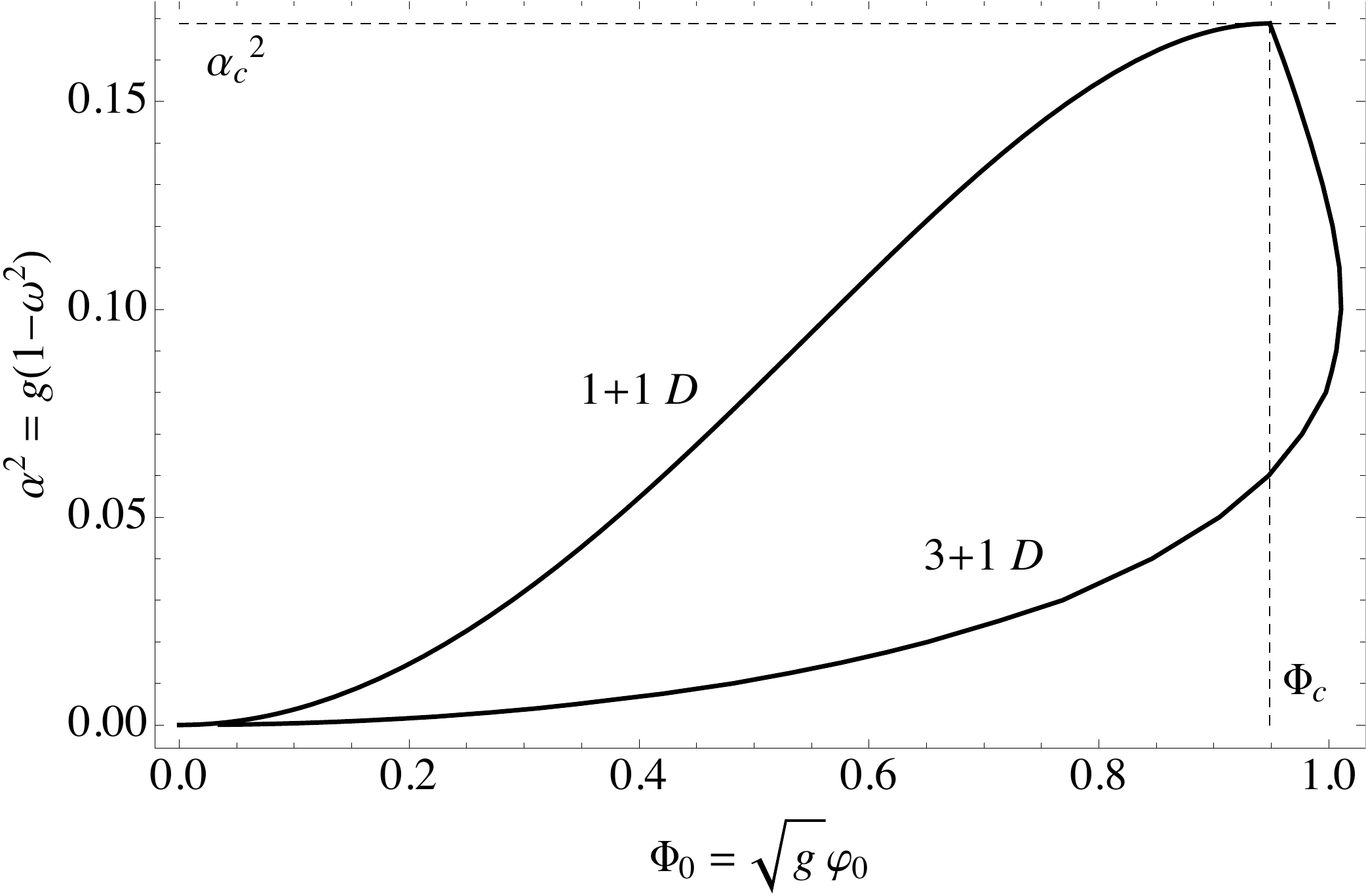} 
   \caption{The above figure shows $\alpha^2$ which characterizes the change in frequency of oscillation due to the nonlinearities in the potential. The critical $\alpha_c=\sqrt{27/160}$ can be obtained from the requirement that the nodeless solution is smooth and localized in space. Note that in $1+1$ dimensions, $\alpha$ is monotonic in $\Phi_0$. This is not the case in $3+1$ dimensions. Frequency is measured in units of the $m$.}
 \label{frequency}
\end{figure}

 \noindent To extract the oscillon profile, we introduce the following change of variables:
\begin{equation}
\begin{aligned}
\label{scaledvar}
& \varphi(t, x)=\frac{1}{\sqrt{g}}\phi\left(\tau,y\right),\\
& t=\omega^{-1}\tau,\\
& y=x/\sqrt{g},\\
\end{aligned}
 \end{equation}
 where $$\omega^2=1-g^{-1}\alpha^2.$$  Here, $\alpha^2$ characterizes the change in frequency due to the nonliear potential. We define $\Phi_0=\phi(0,0)=\sqrt{g}\varphi_0$ and choose $\partial_t\varphi(0,x)=\partial_\tau\phi(0,x)=0$. Note that $\alpha$ and $\Phi_0$ are not independent of each other. Their relationship will be determined from the requirement that the solution is periodic in time, smooth at the origin and vanishing at spatial infinity. With the change of variables (\ref{scaledvar}), and collecting powers of $g$, the equations become
 \begin{equation}
 \partial_\tau^2\phi+\phi+g^{-1}\left[-\alpha^2\partial_\tau^2\phi-\partial_y^2\phi-\phi^3+\phi^5\right]=\mathcal{O}[g^{-3/2}].
 \end{equation}
 Let us consider solutions of the form
 \begin{equation}
 \phi(\tau,y)=\phi_1(\tau,y)+g^{-1}\phi_3(\tau,y)+\hdots
 \end{equation}
Again collecting powers of $g$, we get
\begin{equation}
\begin{aligned}
\label{1dgexp}
&\partial_\tau^2\phi_1+\phi_1=0,\\
&\partial_\tau^2\phi_3+\phi_3=\alpha^2\partial_\tau^2\phi_1+\partial_y^2\phi_1+\phi_1^3-\phi_1^5.\\
\end{aligned}
\end{equation}
The first equation in (\ref{1dgexp}) has a solution of the form
\begin{equation}
\phi_1(\tau,y)=\Phi(y)\cos\tau.
\end{equation}
To determine the profile $\Phi(y)$, we look at the second equation in (\ref{1dgexp}). Substituting $\phi_1(\tau,y)$ into this equation, we get
\begin{equation}
\begin{aligned}
\partial_\tau^2\phi_3+\phi_3=&\left[-\alpha^2\Phi+\partial_y^2\Phi+\frac{3}{4}\Phi^3-\frac{5}{8}\Phi^5\right]\cos\tau,\\
&+\left[\frac{1}{4}\Phi^3-\frac{5}{16}\Phi^5\right]\cos 3\tau-\frac{1}{16}\Phi^5\cos 5\tau.\\
\end{aligned}
\end{equation}
We are looking for solutions that are periodic in $\tau$. The term $[\hdots]\cos\tau$ will lead to a term linearly growing with $\tau$. Hence, we must have
\begin{equation}
\partial_y^2\Phi-\alpha^2\Phi+\frac{3}{4}\Phi^3-\frac{5}{8}\Phi^5=0.
\end{equation}
This equation has a first integral, the ``conserved energy"
\begin{equation}
E_y=\frac{1}{2}\left(\partial_y\Phi\right)^2+\mathcal{U}(\Phi),
\end{equation}
where $\mathcal{U}(\Phi)=-\frac{1}{2}\alpha^2\Phi^2+\frac{3}{16}\Phi^4-\frac{5}{48}\Phi^6$. If we demand spatially localized solutions, we require $E_y=0$. Furthermore, requiring that the profile be smooth at the origin, we must have $\partial_y\Phi(0)=0$. This immediately yields (also see Fig. \ref{frequency})

\begin{equation}
\alpha^2=\frac{3}{8}\Phi_0^2-\frac{5}{24}\Phi_0^4.
\end{equation}
Note that there is a critical value  $\alpha_c=\sqrt{27/160}$ (or $\Phi_c=\sqrt{9/10}$) beyond which localized solutions do not exist.
The profile equation becomes
\begin{equation}
\left(\partial_y\Phi\right)^2=\alpha^2\Phi^2-\frac{3}{8}\Phi^4+\frac{5}{24}\Phi^6.
\end{equation}
Integrating the above equation yields
\begin{equation}
\label{eq:profile1d}
\Phi(y)=\Phi_0\sqrt{\frac{1+u}{1+u\cosh [2\alpha y]}},
\end{equation}
where 
\begin{equation}
\begin{aligned}
&u=\sqrt{1-({\alpha}/{\alpha_c})^2},\\
&\Phi_0=\Phi_c\sqrt{1-u}.
\end{aligned}
\end{equation}
We have introduced the variable $0<u<1$ which simplifies the appearance of the equations and controls the shape of the oscillons through $\alpha$. We will come back to a more detailed analysis of this solution, but first we solve for the second order correction to this solution, $\phi_3$:
\begin{equation}
\begin{aligned}
\partial_\tau^2\phi_3+\phi_3=\left[\frac{1}{4}\Phi^3-\frac{5}{16}\Phi^5\right]\cos 3\tau-\frac{1}{16}\Phi^5\cos 5\tau.\\
\end{aligned}
\end{equation}
The solution with $\phi_3(0,y)=\partial_\tau\phi_3(0,y)=0$ is
\begin{equation}
\begin{aligned}
\phi_3(\tau,y)=\frac{1}{96}\left(3\Phi^3-4\Phi^5\right)\cos\tau+\frac{1}{128}\left(5\Phi^5-4\Phi^3\right)\cos3\tau+\frac{1}{384}\Phi^5\cos5\tau.\\
\end{aligned}
\end{equation}
The full solution becomes 
\begin{equation}
\begin{aligned}
\phi(\tau,y)
=\Phi\cos\tau+\frac{\Phi^3}{24g}\left[\frac{1}{4}\left(3-4\Phi^2\right)\cos\tau-\frac{3}{16}\left(4-5\Phi^2\right)\cos3\tau+\frac{1}{16}\Phi^2\cos5\tau\right].
\end{aligned}
\end{equation}
Note that the corrections to $\phi_1$ are strongly supressed for $g\gg1$. Even for moderately large $g\sim 5$, the factor in the denominator is $\sim 100$, making $\phi_1$ a rather good approximation. From now on we will mainly concern ourselves with $\phi_1$.  

Reverting back to the original variables (\ref{scaledvar}), the solution for $\alpha<\alpha_c$ (equivalently, $\Phi_0<\Phi_c$) is 
\begin{equation}
\label{sol1d}
\varphi(t,x)=\varphi_0\sqrt{\frac{1+u}{1+u\cosh [2\alpha x/\sqrt{g}]}}\cos(\omega t)+\mathcal{O}[g^{-3/2}],
\end{equation}
where 
\begin{equation}
\begin{aligned}
&u=\sqrt{1-({\alpha}/{\alpha_c})^2},\\
&\varphi_0=\frac{\Phi_0}{\sqrt{g}}=\frac{\Phi_c}{\sqrt{g}}\sqrt{1-u},\\
&\omega^2=1-g^{-1}\alpha^2.
\end{aligned}
\end{equation}
Here, $\varphi_0$ is the amplitude of the profile at the origin and scales as $1/\sqrt{g}$. 

\begin{figure}[t] 
   \centering
   \includegraphics[width=6.5in]{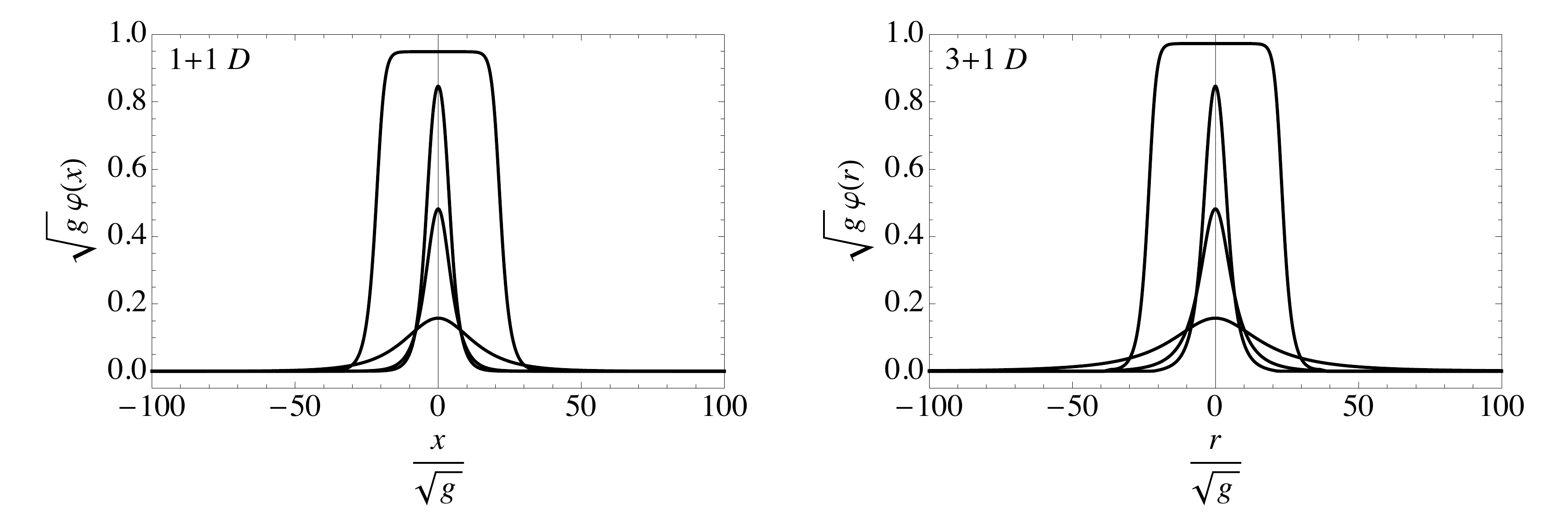} 
   \caption{The above figure shows the spatial profiles of oscillons for different values of the amplitude at the center. For $\Phi_0\ll\Phi_c=\sqrt{9/10}$ we get the usual  sech-like profile, which is consistent with the small amplitude analysis. As $\Phi_0$ approaches $\Phi_c$, the oscillons become wider with surprisingly flat tops. Unlike the $1+1$-dimensional case, in $3+1$ dimensions, we approach the flat-top profiles from above. Distances are measured in units of the $m^{-1}$.}
   \label{profile13d}
\end{figure}

\begin{figure}[t] 
   \centering
   \includegraphics[width=5in]{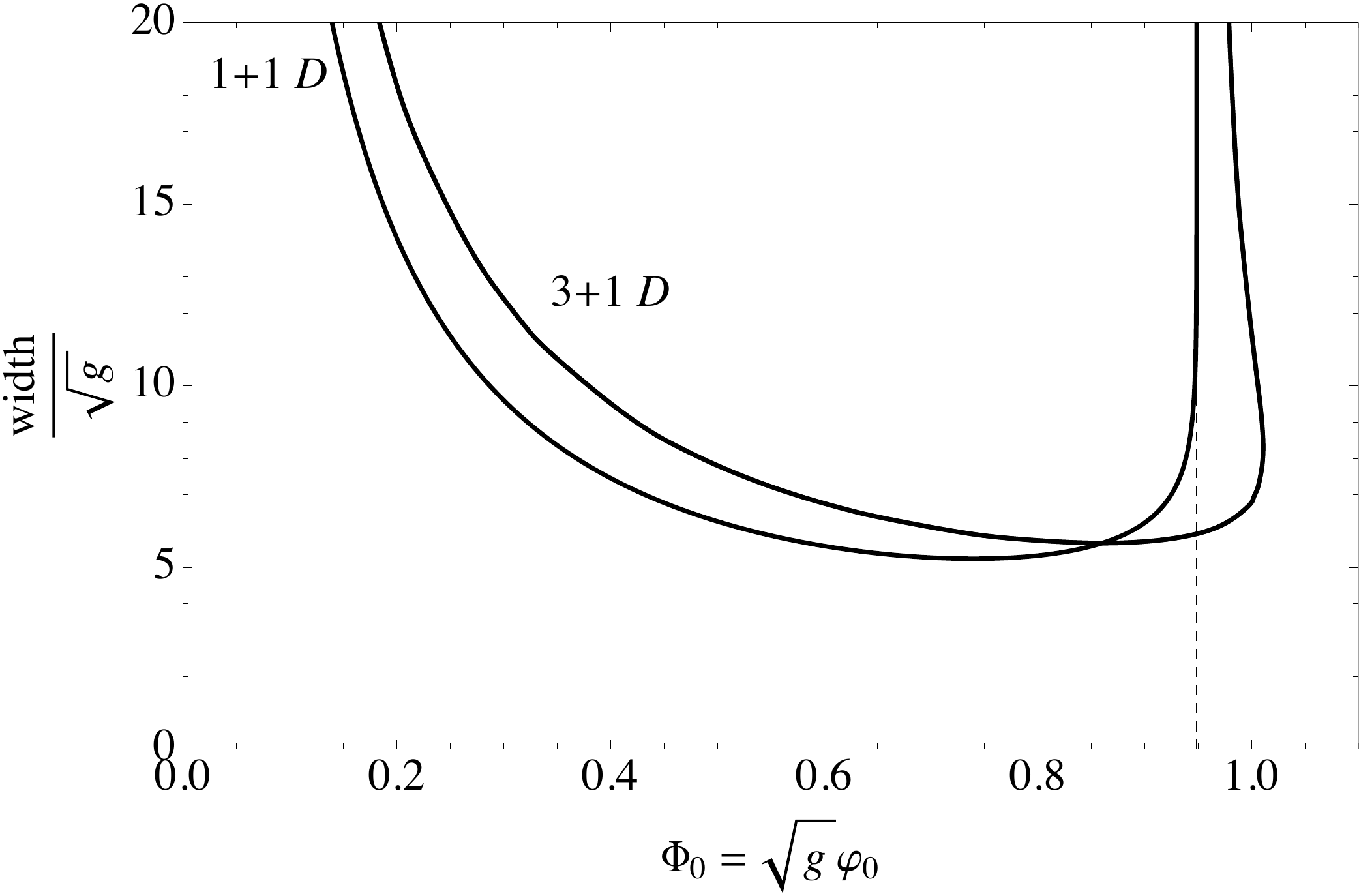} 
   \caption{The above figure shows the nonmonotonic relationship between the width and height of oscillons in $1+1$ and $3+1$ dimensions. Note that as $\Phi_0$ approaches $\Phi_c=\sqrt{9/10}$, the oscillons become wider with flat tops. Unlike the $1+1$-dimensional case, in $3+1$ dimensions, we obtain flat-top profiles when $\Phi_0$ approaches $\Phi_c$ from above.  
   Distances are measured in units of the $m^{-1}$.}
   \label{width13D}
\end{figure}

Let us now investigate the solution for the profile. Figure \ref{profile13d} (top left) shows this solution for different valued of $\alpha$. Notice that as $\alpha$ approaches $\alpha_c$ (equivalently, $\Phi_0\rightarrow \Phi_c$, $u\rightarrow 0$), the oscillon profile begin to deviate from the ``sech" profile and has a flat top. Given this solution, one can derive the width of the oscillon as a function of its height. Defining the width to be the $x$ value where the profile falls by $1/e$ of its maximum

\begin{equation}
\label{width}
x_e=\frac{1}{\varphi_0}\frac{2}{\sqrt{3}}(1+u)^{-1/2}\cosh^{-1}\left[\frac{e^2(1+u)-1}{u}\right].
\end{equation}
As $u\rightarrow 0$ we simply have $x_e\sim 1/\varphi_0$, which is consistent with the small amplitude analysis (see Fig. \ref{width13D}). Meanwhile, $u\rightarrow 1$ yields a spatially uniform solution.

We end this subsection by writing down an expression for the energy of these oscillons:
\begin{equation}
E_{osc}=\varphi_0\frac{4}{\sqrt{3(1-u)}}\tanh^{-1}\left[\sqrt{\frac{1-u}{1+u}}\right]+\mathcal{O}[g^{-3/2}].
\end{equation}
Note that  as $\alpha\rightarrow 0$ ($u\rightarrow 1$) we have $E_{osc}\sim 2\sqrt{{2}/{3}}\varphi_0$ whereas for $\alpha\rightarrow\alpha_c$ ($u\rightarrow 0$), $E_{osc}\rightarrow \infty$. 
In the next subsection we extend our results to $3+1$ dimensions. 

\subsection{Profile and frequency in $3+1$ dimensions}  
\label{profile3d}
  In this section we extend the results of the previous section to a $3+1$-dimensional Minkowski space-time. Although we are unable to obtain an analytic form for the profile, the important qualitative (and some quantitative) aspects of the solutions can still be understood. In particular, we derive a critical amplitude and frequency for which the solution becomes spatially homogeneous and argue that the relationship between the height and width is nonmonotonic. 
  
  The equation of motion (assuming spherical symmetry) is given by
   \begin{equation}
 \partial_t^2{\varphi}-\partial_r^2\varphi-\frac{2}{r}\partial_r\varphi+\varphi-\varphi^3+g\varphi^5=0.
 \end{equation}
 We can follow the same procedure used in the previous subsection to arrive at the equation for the profile
 \begin{equation}
\partial_\rho^2\Phi+\frac{2}{\rho}\partial_\rho\Phi-\alpha^2\Phi+\frac{3}{4}\Phi^3-\frac{5}{8}\Phi^5=0,
\end{equation}
where $\rho=r/\sqrt{g}$.  This is where we first encounter the difficulty associated with three dimensions. We can no longer obtain a first integral due to the $2/\rho(\partial_\rho\Phi)$ term. However, we can still get a bound on $\alpha$ by requiring that the solutions are spatially localized (see \cite{Anderson:1971pt} for an analysis of a similar profile equation in the context of $Q$-balls). It is convenient to define an energy $E_\rho$, which in the absence of the $(2/\rho) \partial_\rho\Phi$ term, is a constant of motion:
\begin{equation}
E_\rho=\frac{1}{2}\left(\partial_\rho\Phi\right)^2+\mathcal{U}(\Phi),
\end{equation}
where $\mathcal{U}(\Phi)=-\frac{1}{2}\alpha^2\Phi^2+\frac{3}{16}\Phi^4-\frac{5}{48}\Phi^6$. With this definition the equation of motion takes on an intuitive form
 \begin{equation}
 \frac{dE_{\rho}}{d\rho}=-\frac{2}{\rho}\left(\partial_\rho\Phi\right)^2.
 \end{equation}
This means that as we move away from $\rho=0$, we move from a higher $E_{\rho}$ trajectory to a lower one. With the requirement that the solution is ``localized" (more specifically, $\Phi\propto \rho^{-1}e^{-\alpha\rho}$ as $\rho\rightarrow\infty$), we need $E_{\rho}\rightarrow 0$ as $\rho\rightarrow\infty$. Requiring that the solution is smooth at $\rho=0$ requires $\partial_\rho\Phi=0$ at $\rho=0$. This implies that for a localized solution we must have $E_\rho\ge0$. Equivalently, $U(\Phi_0)\ge0$, which in turn implies that $\alpha\le\alpha_c=\sqrt{27/160}$. For this critical value $\alpha_c$, we get a special solution [with $\Phi_c(\rho)=\sqrt{9/10}$], which is homogeneous in space. For $0<\alpha<\alpha_c$ we get nonzero spatial derivatives. 

For each $\alpha$ in the range  $0<\alpha<\alpha_c$, only special, discrete values of $\Phi_0(n)$ [$n$ corresponding to the number of nodes]will yield solutions that satisfy our requirement $\Phi\propto \rho^{-1}e^{-\alpha\rho}$ as $\rho\rightarrow\infty$. From these, the $n=0$ ones are the oscillon profiles we are looking for. The numerically obtained profiles are shown on the right in Fig. \ref{profile13d}. 

From figure \ref{width13D}, it is easy to see that the relationship between the heights and widths of the oscillons is nonmonotonic. We know that for $\alpha\ll\alpha_c$ (ie. $\Phi_0\ll\Phi_c$), the usual small amplitude expansion yields solutions that have the property that their widths decrease with increasing amplitude. We also know that for $\alpha=\alpha_c$ ($\Phi_0=\Phi_c$) the width will be infinite. Thus, as in the $1+1$-dimensional scenario, we expect the width to be a nonmonotonic function of the central amplitude. This is indeed what is seen from the numerical solutions of the profile equation as shown in Fig. \ref{width13D}. Note that the width is a multivalued function of the amplitude beyond $\Phi_0=\Phi_c$. Nevertheless, it still approaches the homogeneous solution via the flat-top profiles. The multivalued relationship between $\alpha$ and $\Phi_0$ is shown in Fig. \ref{frequency}.
 \subsection{Radiation}
 
The oscillon solution does not solve the equation of motion exactly. We have ignored terms of $\mathcal{O}[g^{-3/2}]$ as well as outgoing radiation. The problem of calculating the outgoing radiation in the small amplitude limit (not the flat tops) has been addressed in the literature (see \cite{Segur:1987mg,Fodor:2009kf}). Our intention here is to point out that for flat tops, the radiation will still be small.

As shown in \cite{Hertzberg:2009}, the amplitude of the outgoing radiation can be estimated by the amplitude of the Fourier transform of the oscillon at the radiation wave number $k_r\sim \sqrt{8}m$ (also see \cite{Gleiser:2009ys}). For small amplitude oscillons, this is exponentially small $\sim e^{-1/\varphi_0}$. Let us estimate what changes are expected when we move to the flat-top oscillons. 
As we have seen, already our solutions have the form, 
 \begin{equation}
 \varphi(t,x)=\frac{1}{\sqrt{g}}\phi\left(\tau,\frac{x}{\sqrt{g}}\right).
 \end{equation}
 Where the function $\phi(\tau,y)$ is independent of $g$. The Fourier transform of $\varphi$ can be determined from the Fourier transform of $\phi$ using
 \begin{equation}
 \varphi(t,k)=\phi(\tau,\sqrt{g}k).
 \end{equation}
Now $\phi(t,x)$ is determined entirely by  $\alpha$. Hence, for any given $\alpha$, the Fourier transform of $\varphi$ gets narrower as $g$ is increased. Thus, by increasing $g$ we can make the amplitude at the radiating wave number as small as we want. Note that even though the Fourier transform of a flat-top oscillon resembles a "sinc" function, rather than a sech, this is true only for
wave numbers near zero. Since the flat top oscillons are smooth solutions, their Fourier transforms still exhibit a rapid asymptotic decay. The argument in 3+1 dimensions will be similar.


\section{Linear Stability Analysis}
\label{stability}
\begin{figure}[t] 
   \centering
   \includegraphics[width=6.5in]{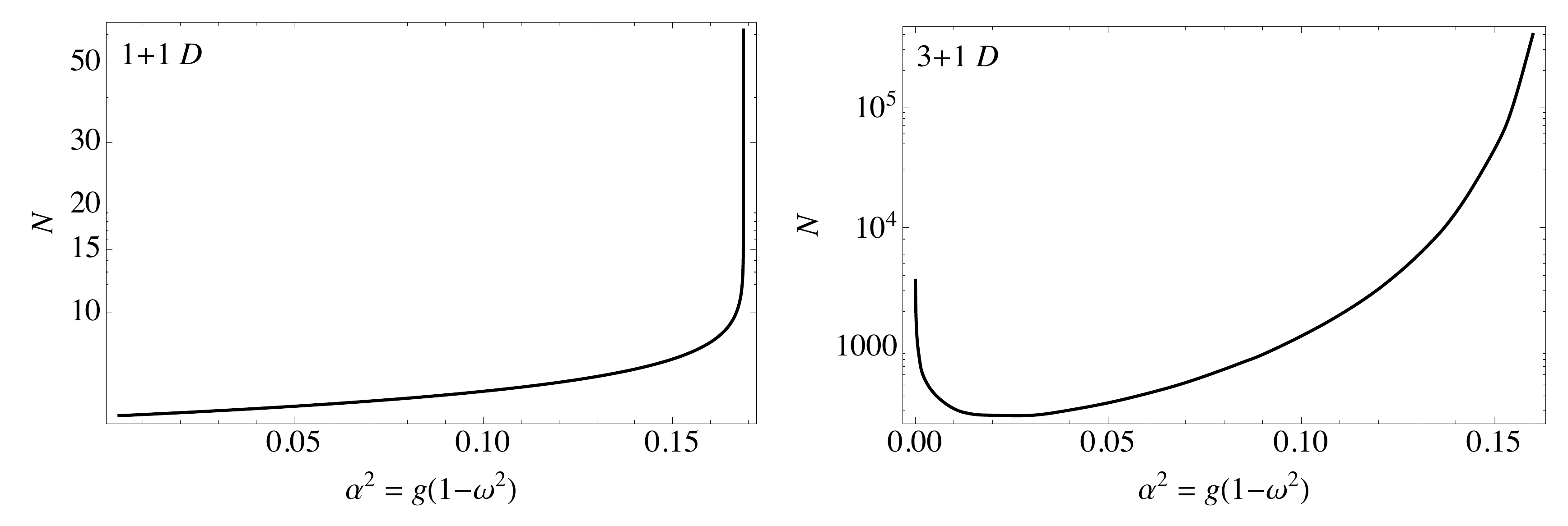} 
   \caption{In the above figure we plot $N=\int \Phi^2 d{\bf x}$ for the oscillons in $1+1$ (left) and $3+1$(right) dimensions. The stability of oscillons is determined by the sign of $dN/d\alpha^2$ where $\alpha^2=g(1-\omega^2)$. Note the important difference between the curves in the two cases. While all oscillons are stable to long wavelength perturbations in $1+1$ dimensions, this is not the case in $3+1$ dimensions. Only those with small frequency (or equivalently, towards the flat-top regime) are robust. Frequency is measured in units of $m$.}
     \label{Nvsalpha}
\end{figure}
In this section we investigate whether oscillons are stable against small, localized perturbations. As discussed in the previous section, the periodic oscillon expansion, formulated in powers of $g^{-1/2}$, fails to solve the governing field equations and must expel radiation.  In our stability analysis we ignore the effects of the exponentially suppressed radiation and focus on perturbing the oscillon profile. The main results of this section are as follows: (i) On the time scale of order $g$, $3+1$-dimensional  oscillons with large amplitudes are robust (their small amplitude counterparts are not) against localized perturbations with spatial variations comparable to the width of the oscillon. (ii) For small wavelength perturbations (compared to the width of of the oscillon), instabilities {\it{could}} exist in discrete, extremely narrow bands in $k$ space. 

We now provide the details essential for reaching the above conclusions. As done previously, we discuss the $1+1$-dimensional case first, and then extend the results to $3+1$ dimensions. Starting with a fixed oscillon profile $\varphi_{osc}$ [see Eq. (\ref{sol1d})], we linearize about the oscillon by an arbitrary function $\chi$. Provided the field $\chi$ remains smaller than $\varphi_{osc}$, the linearized dynamics will approximately describe the perturbation. Let
	\begin{eqnarray}
        \varphi(t, x) &=& \varphi_{osc}(\tau, x) + \delta \chi(t, x),
	\end{eqnarray}
where $\delta \ll g^{-1/2}$ is the amplitude of the perturbation and we keep $\chi \sim O(1)$.  Note that for a linear analysis, $\chi$ must also vanish at infinity so that the perturbation $\delta \chi$ remains smaller than the original oscillon.  Therefore, we restrict our analysis to spatially localized perturbations.  The field $\chi$ then satisfies
	\begin{eqnarray}
	 \label{ChiField}
		\partial_{t}^2 \chi - \partial_{x}^2 \chi + \chi - 3 \varphi_{osc}^2 \chi + 5 g \varphi_{osc}^4 \chi&=& 0.
	\end{eqnarray}
 We now wish to determine if all initial conditions $\chi$ remain bounded, or whether there exists an unstable initial profile $\chi(0, x)$. The $\varphi_{osc}^2(t,x),\varphi_{osc}^4(t,x)$ terms act as periodic forcing functions\footnote{Since $\phi_{osc}$ oscillates in time, the problem is essentially one of parametric resonance stability/instability.   We are really diagonalizing the Floquet matrix  - which in this case would really be an integral operator.}. This periodic forcing, somewhat analogous to pumping ones legs back and forth on a swing, may deposit energy into the field $\chi$ and consequently excite an instability.

 A complete treatment of stability may require one to solve (\ref{ChiField}) for a complete basis of initial conditions. Because of the spatially dependent oscillon solution, a Fourier analysis is difficult. With this in mind, we split the set of initial conditions into two groups. The first with spatial variations comparable to the size of the oscillons and another which varies on much shorter length scales. In the second case we can approximate the oscillon as a spatially constant oscillating background. This allows us to carry out a standard Floquet analysis. Such an analysis reveals the most dangerous instability band at $k\sim\sqrt{3}$, with a width $\Delta k\lesssim g^{-1}\Phi_0^4$.  For large $g$, this becomes extremely narrow. In addition, we expect the time scale of these instabilities to be $\sim g\Phi_0^{-4}$. Nevertheless, one should bear in mind that the slow spatial variation of the oscillon could still be important. 
    
Now, let us look at the case where the perturbations vary on length scales comparable to the width of the oscillon in detail. Note that to leading order, the forcing potential: $\sim\varphi_{osc}^2$ is (i) $\mathcal{O}[g^{-1}]$, (ii) smoothly varying with a natural length $x_e\propto \sqrt{g}$, and (iii) oscillating with period $1$ in the variable $\tau = \omega t$.  The first observation implies we may use perturbation theory and seek an expansion for $\chi$ in inverse powers of $g$:
	\begin{eqnarray}
	\label{chiexp}
		\chi = \chi_0 + g^{-1} \chi_1 + \ldots
	\end{eqnarray}
In the following analysis we shall work out the linear instabilities to first order in $g^{-1}$. However, we must keep in mind that solutions stable to order $g^{-1}$, may in fact develop higher order instabilities over longer time scales. Since we are interested in perturbation with wavelengths comparable to $x_e$, we rescale the length $x = \sqrt{g} y$.  To capture the instability, however, we introduce two times: the original oscillatory time $\tau = \omega t$ and a slow time $T = g^{-1}t$\footnote{One may want to know why $T = g^{-1}t$ provides the important slow time scale. A back of the envelop calculation is as follows - consider a homogeneous background oscillating at the oscillon frequency. A naive perturbation series for $\chi$ in powers of $g^{-1}$ exhibits an oscillating term for $\chi_0$, and then a term which grows linearly in $t$ for $\chi_1$. This means $\chi_1$ becomes the same order as $\chi_0$ when $t g^{-1} \sim 1$. Hence, there is a characteristic slow time $T = t g^{-1}$. If we let $\chi_0$ be a function of both $\tau$ and $T$, and we choose $\chi_0$ correctly, we can ensure that $\chi_1$ remains small for long times. In a similar fashion, if we consider higher order terms in the oscillon expansion, there could be additional instabilities excited over time scales $g^{-2}$, $g^{-3}$ etc. }.  The introduction of $\tau$ follows from our focus on perturbations which oscillate near the oscillon frequency.  In addition, we require a slow time $T$ to capture variations in the perturbation.  Hence, the field $\chi = \chi(\tau, T, y)$ and the derivative $\partial_t$ becomes
    \begin{eqnarray}
    \label{time}
        \partial_t &=& \omega \partial_{\tau} + g^{-1} \partial_T , \label{time1},\\
        \partial_t^2 &=&  \partial_{\tau}^2 + g^{-1} [2 \partial_{T} \partial_{\tau} - \alpha^2 \partial_{\tau}^2]+ \mathcal{O}[g^{-2}]. \label{time2}
    \end{eqnarray}
    Upon substitution of Eqs. (\ref{chiexp}), (\ref{time1}) and (\ref{time2}) in Eq. (\ref{ChiField}) and collecting powers of $g^{-1}$ we obtain
    \begin{eqnarray}
        \partial_{\tau}^2 \chi_0 + \chi_0 &=& 0, \\
        \partial_{\tau}^2 \chi_1 + \chi_1 &=&  -[2 \partial_{T} \partial_{\tau} - \alpha^2 \partial_{\tau}^2 - \partial_y^2 -3 \cos^2 \tau \Phi^2(y) + 5 \cos^4 \tau \Phi^4(y)]\chi_0.
    \end{eqnarray}
    From the zeroth order equation, the most general solution for $\chi_0(\tau, T, y)$ is
    \begin{eqnarray}
        \chi_0(\tau, T, y) = u(T,y)\cos\tau+v(T,y)\sin(\tau)
    \end{eqnarray}
    Here, $u(T, y)$ and $v(T,y)$ are real functions that depend on the slow time $T$ and space $y$.
    Eliminating the secular terms from  the right hand-side of the $\chi_1$ equation, we obtain 
    	\begin{eqnarray} 
		2 \partial_Tu &=& L v, \label{MasterEquations1} \\
		2 \partial_Tv &=& -M u, \label{MasterEquations2}
	\end{eqnarray}
	where the $L$ and $M$ are both Hermitian operators. Explicitly,
	\begin{eqnarray}
		L &=& -\partial_y^2 + \alpha^2 - \frac{3}{4} \Phi^2(y) + \frac{5}{8}\Phi^4(y), \\
		M &=& -\partial_y^2 + \alpha^2 - \frac{9}{4} \Phi^2(y) + \frac{25}{8} \Phi^4(y).
	\end{eqnarray}
	Since Eqs. (\ref{MasterEquations1}) and (\ref{MasterEquations2}) are linear, we can separate variables via $u(T, y) = e^{\frac{1}{2}\Omega T} u(y)$, $v(T, y) = e^{\frac{1}{2}\Omega T} v(y)$:
	\begin{eqnarray}
		\Omega u &=& L v, \\
		\Omega v &=& -M u,
	\end{eqnarray}
	or equivalently
	\begin{eqnarray}
		\Omega^2 u &=& - L M u, \\
	 	\Omega^2 v &=& - M L v. \\
	\end{eqnarray}
	Since both $u$ and $v$ are real fields and $L$ and $M$ are real operators, the eigenvalues\footnote{If they exist.} $\Omega^2$ must also be real.  Hence, all exponents $\Omega$ are either purely real or purely imaginary. Then, oscillon stability is guaranteed when $max (\Omega^2) < 0$, or equivalently when the largest real eigenvalue of $-M L$ is negative.  Determining the largest real eigenvalue of $-ML$ can be done using the analysis performed by Vakhitov and Kolokolov \cite{Vakhitov}.  Specifically, they exploit properties of the operator potentials found in $L$ and $M$ to show that $max (\Omega^2) <0$ if and only if $d N/d \alpha^2 > 0$.  Here, $N$ is the integral over all space:
\begin{eqnarray}
	N &=& \int \Phi^2(y) d y
\end{eqnarray} 
and $\alpha^2=g(1-\omega^2)$. From Fig. \ref{Nvsalpha}, we can see that $dN/d\alpha^2>0$ for all allowed $\alpha$ in $1+1$ dimensions. Thus, $1+1$-dimensional oscillons are stable against small perturbations with long wavelengths. Note that to order $g^{-1}$, $N= 2g^{1/2}E_{osc}$ in $1+1$ dimensions (in $3+1$ dimensions $N= 2g^{-1/2}E_{osc}$.)

The argument of \cite{Vakhitov} holds for dimensions $D = 1, 2, 3$. 
The discussion above carries over to $3+1$ dimensions through the following identifications: $y\rightarrow \rho$ and $\partial_y^2\rightarrow \partial_\rho^2+(2/\rho)\partial_\rho$ and $N = 4\pi\int \Phi^2(\rho) \rho^2d \rho$. The result in $3+1$ dimensions is in sharp contrast with that in $1+1$ dimensions (see Fig. \ref{Nvsalpha}). Unlike the $1+1$-dimensional result, not all oscillons are robust against long wavelength perturbations. Only oscillons with large $\alpha$ (equivalently small frequency or large amplitudes) are robust. This result makes the large amplitude, flat-topped oscillons in $3+1$ dimensions particularly interesting.

In the context of Q-balls, $N$ is proportional to the conserved particle number and plays a role in the stability \cite{Lee:1991ax}. A similar interpretation might be possible here, since to leading order in $g^{-1/2}$, our solution is periodic in time. Finally, we note that the behavior  of $N$ in $1+1$ and $3+1$ dimensions can be understood heuristically. In $1+1$ dimensions, for small $\alpha$, the amplitude of the profile at the origin $\sim\alpha$ whereas the width $\sim 1/\alpha$. Hence, $N\sim\alpha$. For $\alpha\rightarrow \alpha_c$ we have increasingly wide oscillons with amplitudes $\sim \Phi_c$. Hence, $N$ diverges. Now for $3+1$ dimensions, the behavior at $\alpha\rightarrow \alpha_c$ is similar to the $1+1$-dimensional case. However, at small $\alpha$, due to the different spatial volume factor, we get $N\sim 1/\alpha$, therefore implying a nonmonotonic behavior in $N$. 
Based on a numerical analysis in the case of dilatonic scalar fields, it was conjectured in \cite{Fodor:2009kf},  that the stability of oscillon like configurations is related to the slope of the $E_{osc}$ vs amplitude curve. In the large $g$ limit, $E_{osc}\propto N$ and the amplitude $\propto\alpha$. Hence, their conjecture is in agreement with our analytic result.

\section{Including expansion}  
\label{expansion}

\noindent In this section we consider the effects of expansion on the lifetimes and shapes of oscillons.  We closely follow the procedure provided in \cite{Farhi:2007wj} for the small amplitude oscillons. Here, applying their procedure is somewhat subtle since in the limit $g\gg1$, oscillons tend to be very wide ($\propto\sqrt{g}$), and the width grows without bound when $\alpha\rightarrow 0,\alpha_c$. Consequently, in these regimes it is easier to break up the oscillons due to Hubble horizon effects. Nevertheless we construct approximate solutions when the oscillon width is small compared to the Hubble horizon. 
\subsection{Including expansion in $1+1$ dimensions} 
As before, we begin with $1+1$ dimensions and generalize to $3+1$ dimensions.
We will work in static deSitter co-ordinates where the metric is given by
\begin{equation}
ds^2=-(1-x^2H^2)dt^2+(1-x^2H^2)^{-1}dx^2.
\end{equation}
Here, $H$ is a constant Hubble parameter\footnote{The assumption of $H$ being constant is for simplicity. The analysis carries over to a time dependent $H$ as long as the frequency of oscillation $\omega\gg H$.}. In these co-ordinates, the equation of motion becomes
\begin{equation}
\label{eq:EOMexp1D}
(1-x^2H^2)^{-1}\partial_t^2\varphi+2xH^2\partial_x\varphi-(1-x^2H^2)\partial_x^2\varphi=-V'(\varphi),
\end{equation}
where $(xH)<1$. We will assume that $H\ll1$ and that $H=\bar{H}/g$ where $\bar{H}$ is a small number. The effects of expansion can be ignored when $x\ll H^{-1}$. For oscillons with widths satisfying $x_e(\alpha)\ll H^{-1}$, the solution to the above equation is well approximated by the Minkowski space solution. However, in the tail of the oscillon profile we cannot ignore the effects of expansion. Nevertheless, taking advantage of the exponential decay of the profile in the tails, we can linearize Eq. (\ref{eq:EOMexp1D}) and obtain a solution using the WKB approximation. 

\begin{figure}[t] 
   \centering
   \includegraphics[width=6.5in]{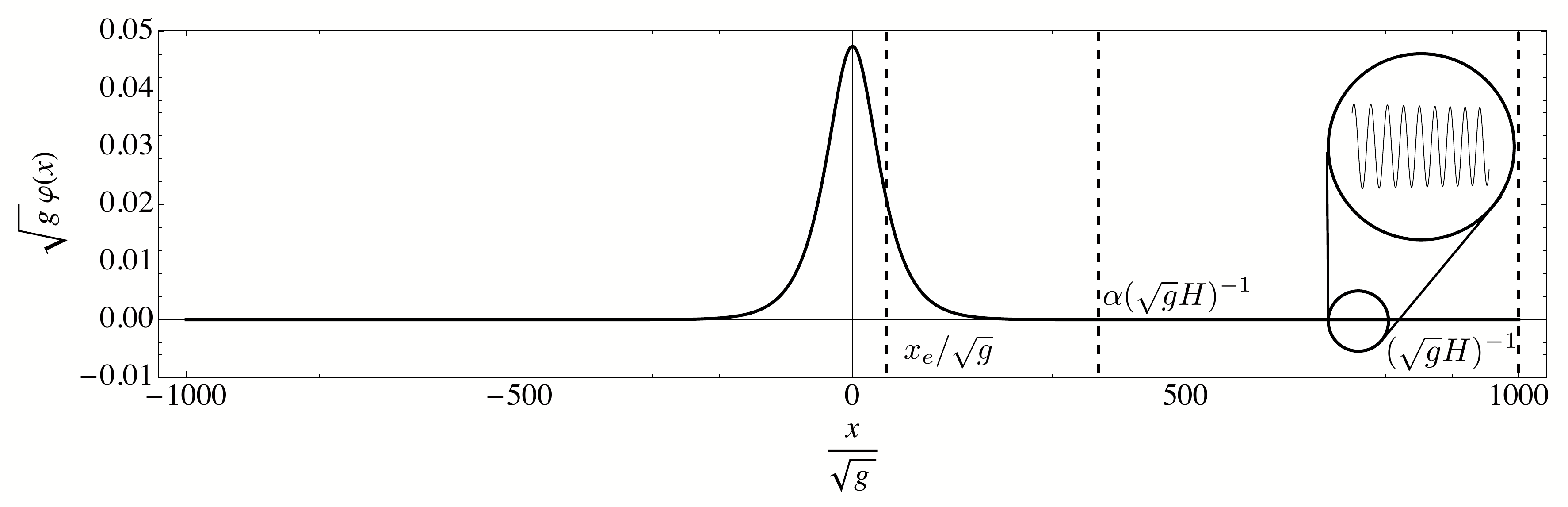} 
   \caption{In an expanding background, if the width is small compared to $H^{-1}$, the flat space solution is adequate for distances much less than $\alpha (\sqrt{g}H)^{-1}$. For distances larger than this, but still smaller than $H^{-1}$, the oscillon feels the expansion, and loses energy in the form of outgoing waves (see inset in Fig. \ref{HOsc}). Distance is  measured in units of the $m^{-1}$.} 
   \label{HOsc}
\end{figure}

We carry out the change of space-time variables and redefinition of the field as was done in the nonexpanding case, Eq. (\ref{scaledvar}). Again collecting powers of $g$, we get

\begin{equation}
\begin{aligned}
&\partial_\tau^2\phi_1+\phi_1=0,\label{gexp1}\\
&\partial_\tau^2\phi_3+\phi_3=\{-\alpha^2+y^2\bar{H}^2\}\partial_\tau^2\phi_1-\partial_y^2\phi_1-\phi_1^3+\phi_1^5. \\
\end{aligned}
\end{equation}
In the case of the Minkowski background, we chose an initial condition $\partial_t\varphi(0,x)=0$, which picked out one of the two linearly independent solutions of the first equation in (\ref{gexp1}). However, in the expanding universe we need to keep the general solution
\begin{equation}
\phi_1(\tau,y)=\frac{\Phi(y)}{2}e^{-i\tau}+c.c,
\end{equation}
where $\Phi(y)$ can be complex and $c.c$ stands for complex conjugate.
The ``profile" equation is given by
\begin{equation}
\left\{\alpha^2-(y\bar{H})^2\right\}\Phi-\partial_y^2\Phi-\frac{3}{4}|\Phi|^2\Phi+\frac{5}{8}|\Phi|^4\Phi=0
\end{equation}
and includes the effect of expansion through the $(y\bar{H})^2$ term. We now analyze different regimes as seen in Fig. \ref{HOsc}. For $(y\bar{H})^2\ll \alpha^2$, the equation admits solutions identical to the nonexpanding case [see Eq. (\ref{sol1d})]. 
 In the region $y_e(\alpha)\ll y \ll\alpha\bar{H}^{-1}$, where $y_e(\alpha)$ is the approximate width of the oscillon [Eq. (\ref{width})], the profile has the form
\begin{equation}
\label{eq:decay1d}
\Phi(y)\approx \Phi_0\sqrt{\frac{2(1+u)}{u}}\exp\left[-\alpha y\right]\hspace{1.5cm}y_e(\alpha)\ll y \ll \alpha\bar{H}^{-1}.
\end{equation}
Since this is an exponentially decaying solution, we can ignore the nonlinear terms in the potential when $y\gg y_e(\alpha)$:
\begin{equation}
\partial_y^2\Phi+\left\{(y\bar{H})^2-\alpha^2\right\}\Phi\approx0\hspace{1.5cm}y_e(\alpha)\ll y.
\end{equation}
For $y>\alpha \bar{H}^{-1}$, the above equation has a WKB solution\footnote{assuming the WKB condition $\bar{H}/\alpha^2\ll1$ is satisfied} in the form of an outgoing wave: 
\begin{equation}
\Phi(y)\approx\Phi_0\sqrt{\frac{2\alpha(1+u)}{u\bar{H}y}}\exp\left[{-\frac{\pi\alpha^2}{4\bar{H}}+\frac{i}{2}\bar{H}y^2}\right].\\
\end{equation}

The amplitude of the outgoing wave was chosen using the WKB connection formula to match the oscillon profile in Eq. (\ref{eq:decay1d}). In terms of the original variables, we obtain
\begin{equation}
\label{solexp1d}
\begin{aligned}
&\varphi(t,x)=\varphi_0\sqrt{\frac{1+u}{1+u\cosh\left[ 2\alpha x/\sqrt{g}\right]}}\cos(\omega t)+\mathcal{O}[g^{-3/2}]\hspace{1.5cm} |x|\ll \alpha(\sqrt{g}H)^{-1},\\
&\varphi(t,x)=\varphi_0\sqrt{\frac{2(1+u)\alpha}{g^{1/2}H|x|u}}e^{-\frac{\pi\alpha^2}{4gH}}\cos\left[\omega t-\frac{1}{2}Hx^2\right]\hspace{1.5cm}   \alpha(\sqrt{g}H)^{-1}\ll|x|< H^{-1},\\
\end{aligned}
\end{equation}
where
\begin{equation}
\begin{aligned}
&u=\sqrt{1-({\alpha}/{\alpha_c})^2},\\
&\varphi_0=\frac{\Phi_c}{\sqrt{g}}\sqrt{1-u},\\
&\omega^2=1-g^{-1}\alpha^2.
\end{aligned}
\end{equation}
Our solution matches that of \cite{Farhi:2007wj} in the limit $\alpha\ll\alpha_c$. However, as $\alpha$ gets larger the coefficient in front of the traveling wave captures the effects of the flat-top solutions. We will return to the above solution when we discuss the rate of energy loss by oscillons after considering the effects of expansion in $3+1$ dimensions. 

We end the section by reminding ourselves of the assumptions required for this solution to be valid: (i) $g\gg1$, (ii)$H<\mathcal{O}[g^{-1}]$, and (iii)
\begin{equation}
{x_e(\alpha)}\ll {\alpha}({\sqrt{g}H})^{-1}.
\end{equation}
For any $H$, the solution is not valid when $\alpha\rightarrow0$ or $\alpha_c$. Also note that 
 for a given $H\ll1$, there always exists a $g$, which violates condition (iii) for all allowed $\alpha$. 

\subsection{Including expansion in $3+1$ dimensions}
Now, let us include the effects of expansion for the $3+1$-dimensional cases. The metric in the static deSitter co-ordinates (assuming spherical symmetry) is given by
\begin{equation}
ds^2=-(1-r^2H^2)dt^2+(1-r^2H^2)^{-1}dr^2+r^2d\Omega^2.
\end{equation}
Following a procedure similar to the one we laid out for the $1+1$-dimensional case, we get the profile equation:
\begin{equation}
\left\{\alpha^2-(\rho\bar{H})^2\right\}\Phi-\partial_\rho^2\Phi-\frac{2}{\rho}\partial_\rho\Phi-\frac{3}{4}|\Phi|^2\Phi+\frac{5}{8}|\Phi|^4\Phi=0,
\end{equation}
where $\rho=r/\sqrt{g}$. The effect of expansion is included through the $(\rho\bar{H})^2$ term. 
For a given $\alpha$, let the approximate width of the oscillon be $\rho_e(\alpha)$. In the region $\rho_e(\alpha)\ll \rho \ll \alpha\bar{H}^{-1}$, the profile has the form
\begin{equation}
\Phi(\rho)\approx f(\alpha)\frac{1}{\rho}\exp\left[-\alpha\rho\right]\hspace{1.5cm}\rho_e(\Phi_0)\ll \rho \ll \alpha\bar{H}^{-1}
\end{equation}
The lack of an analytic solution, prevents us from specifying $f(\alpha)$. Reverting back to the original variables, the solution in the spatially oscillatory regime is given by
\begin{equation}
\begin{aligned}
\label{exp3Dsol}
&\varphi(t,r)=(g^{-1/2}\alpha)^{1/2}\frac{f(\alpha)}{\sqrt{Hr^3}}e^{-\frac{\pi \alpha^2}{4gH}}\cos\left[\omega t-\frac{1}{2}Hr^2\right]\hspace{1.5cm}  \alpha (\sqrt{g}H)^{-1}\ll r< H^{-1},\\
\end{aligned}
\end{equation}
where $\omega^2=1-g^{-1}\alpha^2$.

 \subsection{Energy loss due to expansion}
In this subsection, we discuss the energy loss suffered by oscillons due to the expanding background. As before, we start with the $1+1$-dimensional scenario and then generalize to $3+1$ dimensions. The energy lost by an oscillon whose width is small compared to $H^{-1}$ is given by
\begin{equation}
\label{Eloss1d}
\frac{dE_{osc}}{dt}=-T^x_t |_{-X}^{X},
\end{equation}
where $T^\mu_\nu$ is the energy momentum tensor of the scalar field. We have ignored the dependence of the metric on $x$. We take $X$ to be in the region sufficiently far away from the center. More explicitly, we consider X such that 
\begin{equation}
 \alpha (\sqrt{g}H)^{-1}\ll|X|< H^{-1}.
\end{equation}
 Using Eq. (\ref{solexp1d}) in Eq. (\ref{Eloss1d}), we get (to leading order in $HX^2$ and $g^{-1/2}$)
\begin{equation}
\begin{aligned}
\frac{dE_{osc}}{dt}\approx -\varphi_0^3\left(\sqrt{\frac{3}{4}}\frac{(1+u)^{3/2}}{u}\right) \exp[-\pi\alpha^2/2gH]
\end{aligned}
\end{equation}
\begin{figure}[t] 
   \centering
   \includegraphics[width=6.5in]{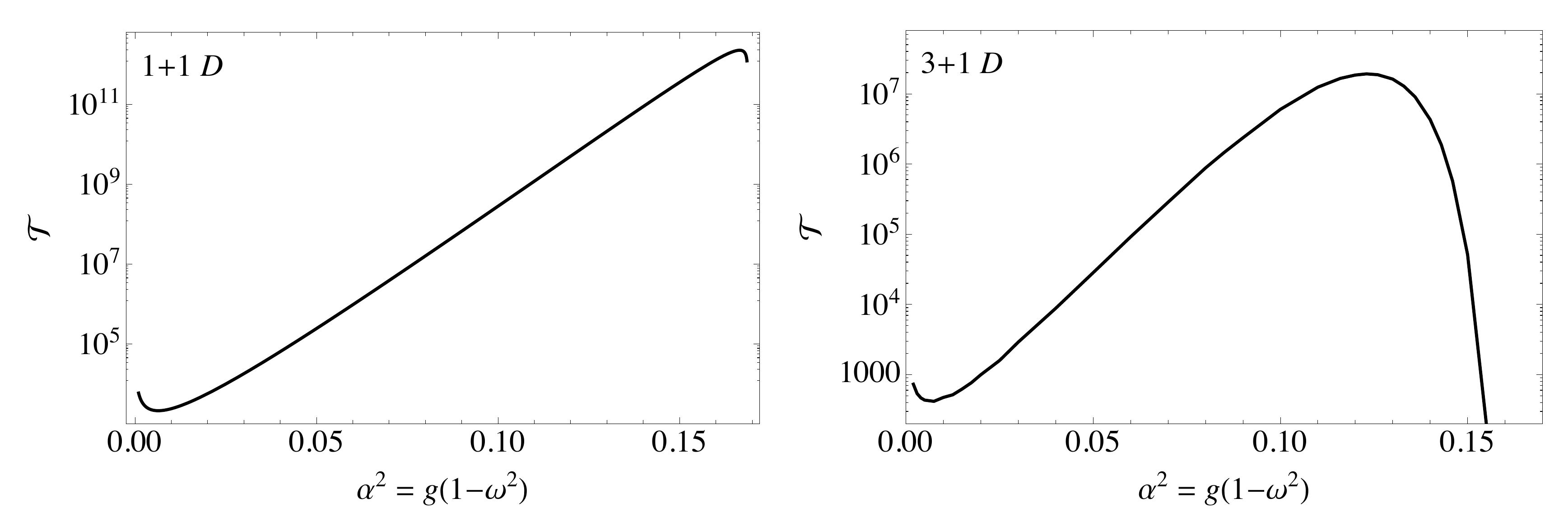} 
   \caption{The above plot shows the time it takes for an oscillon to lose most of its energy due to the expanding background. The plot assumes an $H=10^{-3}$ and $g=10$. The lifetime is sensitive to the combination $\sim g\exp[\alpha^2/(gH)]$. Our result is only valid when $\alpha$ is not too close to $0$ or $\alpha_c=\sqrt{27/160}$, since for these value our assumption $r_e\ll\alpha(\sqrt{g}H)^{-1}$ breaks down. Note that the lifetime should be interpreted qualitatively, since we are not allowing for instabilities that lead to the oscillon's abrupt disintegration. Time is measured in units of $m^{-1}$.}
   \label{lifetime}
\end{figure}
In \cite{Farhi:2007wj}, a similar expression was provided in the limit where $u\rightarrow 1$. Oscillons are known to lose energy very slowly for a long time and then suddenly disintegrate. Our calculation cannot capture this disintegration. We therefore take the following as an upper bound on the lifetime of an oscillon (approximated by the time it take for it to lose most of its energy due to expansion effects)
\begin{equation}
\mathcal{T}\lesssim{E_{osc}}\left|\frac{dE_{osc}}{dt}\right|^{-1}=\frac {1} {\varphi_0^2}\left(\frac{8u\tanh^{-1}\left[\sqrt{\frac{1-u}{1+u}}\right]} {3\sqrt{1-u^2}(1 + u)}\right) \exp[\pi\alpha^2/2gH]
\end{equation}
where
\begin{equation}
\begin{aligned}
&u=\sqrt{1-({\alpha}/{\alpha_c})^2},\\
&\varphi_0=\frac{\Phi_c}{\sqrt{g}}\sqrt{1-u},\\
\end{aligned}
\end{equation}
The terms in $(\hdots)$ are corrections for the deviation from the sech profile. We plot the lifetime as a function of the $\alpha^2$ in Fig. \ref{lifetime}. We should not trust this curve in the limit  $\alpha\rightarrow 0,\alpha_c$, since our assumption of $x_e(\alpha)\ll(\sqrt{g}H)^{-1}$ is broken here. Note the scaling with $g$: $\mathcal{T}\sim g\exp[\alpha^2/(gH)]$. 

A numerical investigation of oscillon lifetimes in a $1+1$-dimensional expanding background was carried out in \cite{Graham:2006xs}: however, no analytic calculation was provided for the lifetime. Our analytic results seem to be in good qualitative agreement with their paper. Note that $E_{osc}\sim g^{-1/2}\alpha^2$ for small $\alpha$, which tells us that the lifetime $\mathcal{T}\sim \exp[E_{osc}/g^{3/2}H]$, is in agreement with the empirical formula inferred in their paper (also see Fig. 2 in \cite{Graham:2006xs}).

A similar calculation can be carried out in $3+1$ dimensions. The rate of energy loss is then given by (to leading order in $g^{-1/2}$)
\begin{equation}
\begin{aligned}
\frac{dE_{osc}}{dt}\approx -\frac{4\pi}{\sqrt{g}} f^2(\alpha)\alpha\, e^{-\pi\alpha^2/(2gH)}.
\end{aligned}
\end{equation}
Here, we have used Eq. (\ref{exp3Dsol}).
The lifetime is then given by
\begin{equation}
\mathcal{T}\lesssim\frac {N(\alpha)} {8\pi f^2(\alpha)\alpha} g e^{\pi\alpha^2/(2gH)}.
\end{equation}
Note that $2E_{osc}(\alpha)=g^{1/2} N(\alpha)$ up to leading order in powers of $g$. We plot the approximate lifetime of an oscillon in a $3+1$-dimensional scenario in Fig. \ref{lifetime}. The lifetime is maximized at $\alpha<\alpha_c$. In plotting the lifetime, the numerically obtained $N(\alpha)$ and $f(\alpha)$ had errors of about 1 percent. Again, we stress that one should think about the above curve somewhat qualitatively since higher order instabilities are ignored. It would be interesting to numerically check the above analysis using a full $3+1$-dimensional evolution.
\section{Conclusions}
\label{con}
In this paper, we pointed out what is required for scalar field potentials to support oscillons. 
We found that the spatial profiles can be very different from a Gaussian, an ansatz often made in the literature. In particular, we derived the nonmonotonic relationship between the height and the width of the oscillons, and discussed the importance of this feature for the stability of oscillons. We showed that the flat-topped oscillons are more stable in three dimensions to long wavelength perturbations as compared to their usual Gaussian counterparts. To the best of our knowledge, this had not been done previously in the literature in the context of oscillons. Oscillons could have important applications in cosmology, especially in the early Universe.  With this in mind, we discussed the changes in the profile, loss of energy from these oscillons due to expansion, and estimated their lifetimes. We provided analytic results for the $1+1$-dimensional scenario, and extended analytically where possible to $3+1$ dimensions and  numerically otherwise. 

A number of questions related to this work require further investigation. Our expressions for lifetime and arguments for stability, especially in $3+1$ dimensions in the flat-top regimes should be checked with a detailed numerical investigation. The question of the possible small wavelength, narrow band instability needs to be resolved rigorously. Recently, \cite{Fodor:2009kg} discussed oscillons in the presence of gravity (an oscillaton). It would be interesting to revisit this problem in the context of our large energy, flat-top oscillons. A study of oscillons emerging from (p)reheating-\cite{Kofman:1997yn,Shtanov:1994ce} like initial conditions in the early Universe is currently in progress.

\section{Acknowledgments}
We would like to thank Eddie Farhi, Alan Guth,  Mark Hertzberg, Ruben Rosales, and Evangelos Sfakianakis, Mark Mezei, Bob Wagoner and Richard Easther for many stimulating discussions, insightful comments, and help though the various stages of the project. MA would particularly like to thank Alan Guth for introducing him to oscillons. MA is supported by a Pappalardo Fellowship at MIT. DS is supported by a Post Graduate Scholarship from the Natural Sciences and Engineering Research Council (NSERC) of Canada. 


\end{document}